\documentclass{aa}  
\usepackage{graphicx}
\usepackage{txfonts}
\usepackage{lipsum}
\usepackage{subcaption}
\usepackage{lscape}
\usepackage{placeins}
\usepackage[colorlinks=true, citecolor=blue, linkcolor=black, urlcolor=black]{hyperref}
\usepackage{orcidlink}
\usepackage{bm}
\usepackage{enumitem}

\begin{document}
\title{3D Radiative Transfer of Lyman-series Lines with \texttt{SKIRT}}

\author{
N. Sameshima\inst{1,2}\email{naoto@ac.jaxa.jp}$^{\orcidlink{0009-0005-6772-5529}}$\fnmsep\thanks{naoto@ac.jaxa.jp}
\and 
A. Lauwers\inst{3}\email{arno.lauwers@ugent.be}$^{\orcidlink{0009-0003-9692-9382}}$
\and 
B. Vander Meulen\inst{3,4,5}\email{bert.vandermeulen@esa.int}$^{\orcidlink{0000-0002-5488-1961}}$
\and 
\\
M. Tsujimoto\inst{1,2}\email{tsujimoto.masahiro@jaxa.jp}$^{\orcidlink{0000-0002-9184-5556}}$\fnmsep\thanks{tsujimoto.masahiro@jaxa.jp}
\and
L.-Y. Gu\inst{5}\email{l.gu@sron.nl}$^{\orcidlink{0000-0001-9911-7038}}$
\and 
M. Baes\inst{3}\email{maarten.baes@ugent.be}$^{\orcidlink{0000-0002-3930-2757}}$
\and 
P.~Camps\inst{3}\email{peter.camps@me.com}$^{\orcidlink{0000-0002-4479-4119}}$
}
\institute{
Institute of Space and Astronautical Science, Japan Aerospace Exploration Agency, 3-1-1 Yoshino-dai, Chuo-ku, Sagamihara, Kanagawa 252-5210, Japan
\and Department of Astronomy, Graduate School of Science, The University of Tokyo, 7-3-1 Hongo, Bunkyo-ku, Tokyo 113-0033, Japan
\and Department of Physics and Astronomy, Universiteit Gent, Proeftuinstraat 86 N3, B-9000 Ghent, Belgium
\and European Space Agency, European Space Research and Technology Centre, Keplerlaan 1, 2201 AZ Noordwijk, The Netherlands
\and SRON Netherlands Institute for Space Research, Niels Bohrweg 4, 2333 CA Leiden, The
Netherlands
}
\date{\today}

\abstract
{High-resolution X-ray spectroscopy and polarimetry provided by XRISM and IXPE offer new
diagnostics of the geometry and kinematics of photo-ionised plasmas around compact
objects. Interpreting reprocessed X-ray emission in such systems requires full
three-dimensional radiative transfer (3D RT) including photon-ion interactions.}
{We extend the Monte Carlo (MC) RT code \texttt{SKIRT} by implementing the Lyman-series
lines of H-like ions, enabling self-consistent modelling of resonance scattering,
radiative recombination, and polarisation of these lines in X-ray photo-ionised plasmas.}
{We implemented Lyman-series transitions (up to $n=10$) for ions with $Z=$1--30,
including fine-structure splitting and linear polarisation in resonance scattering. Two
channels for the production of the Lyman-series lines (resonance scattering and radiative
recombination) are considered. The microphysics (cross-sections, branching ratios, and
redistribution functions) are validated against analytical solutions and compared with
the 1D RT code \texttt{Cloudy} based on the two-stream solver. We further demonstrate
the capabilities of the code using idealised geometrical setups and a 3D geometry of a low-mass X-ray
binary.}
{The implementation reproduces analytical expectations and shows good agreement with
\texttt{Cloudy}. The \texttt{SKIRT} simulations naturally capture RT effects such as P
Cygni profiles and line-profile distortion in optically thick media, which are
inaccessible to the conventional 1D RT codes commonly used in X-rays. In 3D geometries,
we find that anisotropic illumination and velocity fields significantly modify the Lyman
series line ratios and profiles, all of which are observable with XRISM.}
{The extended version of \texttt{SKIRT} provides a powerful framework for interpreting X-ray line
spectra and polarisation from photo-ionised plasmas. It is particularly suited for
constraining the geometry and velocity structure in the vicinity of compact objects in
the XRISM and IXPE era.}
\maketitle
\nolinenumbers

\section{Introduction}\label{s1}
Compact objects, such as black holes (BHs) and neutron stars (NSs), are sites of mass
accretion with deep gravitational potentials. A large fraction of the liberated
gravitational energy is emitted as X-ray radiation, which can drive matter outward in
the form of radiatively driven winds. Characterising the inflow-outflow structure
around compact objects is key to understanding the circulation of matter and energy in
the Universe. Direct imaging of such structures is nearly impossible with current
technology; however, spectroscopy and polarimetry provide a wealth of information. This
approach has become even more powerful in recent years with the launch of the X-ray
Imaging and Spectroscopy Mission (XRISM; \citealt{T2025c}) and the Imaging X-ray
Polarimetry Explorer (IXPE; \citealt{W2022}), which deliver X-ray spectroscopic and
polarimetric data of unprecedented quality.

The observed radiation can be decomposed into transmitted and reprocessed components. The
transmitted component is useful for probing matter along the line of sight through absorption
features imprinted on the otherwise featureless spectra of compact objects. The
reprocessed component carries information about the surrounding material through
absorption/re-emission and scattering. Numerical radiative transfer (RT) calculations
are essential for interpreting the observed radiation, which consists of these two
components that exhibit features produced by multiple physical processes.  Among various
numerical RT codes, one-dimensional (1D) codes based on the two-stream solver are
commonly used in the X-rays. In contrast, the Monte Carlo (MC) solver has an advantage
in its flexibility in modelling three-dimensional (3D) structures
\citep{S2013a,N2019}. This advantage is especially important when dealing with
reprocessing processes, which must be considered within fully 3D geometries.

\texttt{SKIRT} \citep{C2015,B2015,C2020} is a state-of-the-art MC-RT code that features a
wealth of built-in geometries, radiation sources, transfer media characterisations,
spatial grids, and detectors, in addition to various mechanisms for importing models
generated by hydrodynamical simulations. The code is publicly available as open-source 
software\footnote{\url{https://github.com/SKIRT/SKIRT9}}, with detailed 
documentation and various tutorials\footnote{\url{https://skirt.ugent.be}}.
For interpreting observational data, the most
compelling reason to adopt \texttt{SKIRT} is its implementation of a suite of modern
acceleration techniques, including peel-off, forced scattering, continuous absorption,
composite biasing, and explicit absorption \citep{B2011,B2016,B2022a}. Without these
techniques, MC-RT codes would suffer from severe MC noise 
when applied to recent high-resolution data at reasonable computational cost.

\texttt{SKIRT} was initially developed to study interstellar dust grains \citep{B2003}
in the sub-mm to ultraviolet bands. Additional physical processes were incorporated over
time to enhance its versatility and utility across wider wavebands, such as atomic and
molecular excitation in the millimetre band \citep{M2023c}. Atomic processes in the
X-ray band were introduced recently, beginning with the scattering by free and bound
electrons and dust grains, as well as photoelectric absorption and fluorescence by cold
neutral matter \citep{M2023b,M2024}. However, processes involving ions have not yet been
implemented with a notable exception of the provisional work by \cite{T2024}. These
processes are most critical to interpret high-resolution X-ray spectra of compact
objects rich in numerous features arising from highly ionised (H-like and He-like)
ions. These features are spectrally resolved with XRISM and contain a wealth of
diagnostic information when properly interpreted through RT calculations.

We have now embarked on implementing atomic processes involving these highly ionised 
ions in \texttt{SKIRT}. In this paper, we present the simplest case as a first step, 
in which we implement processes responsible for the Lyman-series lines of H-like ions.

The outline of this paper is as follows. We start with the scope (Sect.~\ref{s2-1}),
physical processes (Sect.~\ref{s2-2}), and formalism (Sect.~\ref{s2-3}) in Sect.~\ref{s2}. We
consider two channels: resonance scattering (RS) and radiative recombination (RR). The
implementation is presented in Sect.~\ref{s3} separately for RS (Sect.~\ref{s3-1}) and RR
(Sect.~\ref{s3-2}), followed by verification tests in Sect.~\ref{s4}. In Sect.~\ref{s5}, we
demonstrate the features of the new implementation in a representative
application.

\section{Physics}\label{s2}
\subsection{Scope}\label{s2-1}
In this implementation, we focus on a low-density, photo-ionised plasma surrounding
compact objects. In such plasmas, radiative processes dominate over collisional
processes in determining the charge and level populations. They are in non-local
thermodynamic equilibrium (NLTE) and RT calculations are needed to solve these populations
and synthesise X-ray spectra. Photo-ionised plasmas are characterised by the ionisation
parameter $\xi \equiv L_\mathrm{X}/n_\text{H}r^{2}$ \citep{T1973}, in which $L_\mathrm{X}$ is
the incident X-ray luminosity in the 1--1000~Ryd range, $n_\text{H}$ is the local density of the
plasma, and $r$ is the distance from the X-ray source to the plasma.

We further focus on the principal quantum number $n \rightarrow 1$ transition lines of
H-like ions (Lyman-series lines), which are among the most conspicuous spectral features
observed with XRISM. For the Ly$\alpha$ line of H, the resonance scattering has already
been implemented in \texttt{SKIRT} \citep{C2021}. Based on this, we make the following
extensions: (1) we apply to other elements up to the atomic number $Z=30$, (2) we also
include $n>1$ levels to account for higher Lyman series, (3) we distinguish the
fine-structure levels of Ly$\alpha_1$ and Ly$\alpha_2$, and (4) we include linear (but
not circular) polarisation. These extensions are motivated by recent observations with
XRISM, in which the fine structure levels of Fe Ly$\alpha_1$ and Ly$\alpha_2$ are
resolved \citep{G2025a} and higher Lyman-series lines were detected up to Ly$\eta$
($n=8$) \citep{A2025a}. The line ratios of Ly$\alpha_1$/Ly$\alpha_2$ and the Lyman
decrement (Ly$\beta$/Ly$\alpha$) are useful diagnostics of plasma around compact objects
\citep{T2025a}, which are altered by RT effects including linear polarisation.

\subsection{Atomic processes}\label{s2-2}
\begin{figure}
 \begin{center}
  \includegraphics[width=1.0\columnwidth,clip]{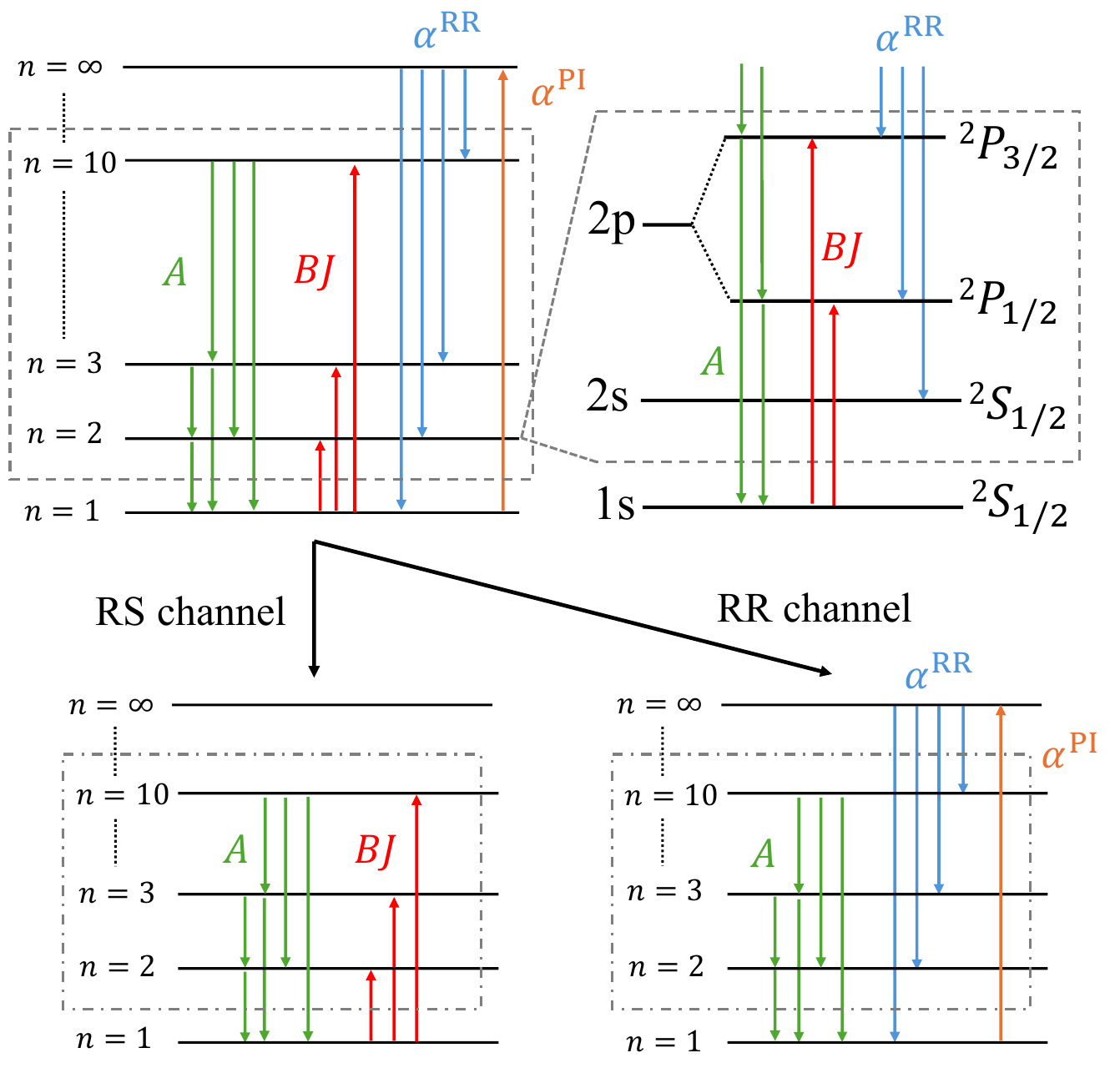}
 \end{center}
 \caption{Levels and atomic processes considered in the present implementation. $A$ and
 $B$ are the Einstein coefficients, $J$ is the radiation field strength and $BJ$
 represents the radiative excitation rate. RS, RR and
 PI denote resonance scattering, radiative recombination and photo-ionisation,
 respectively. Other symbols are defined in the main text.}
 \label{f01}
\end{figure}

\begin{figure}
 \begin{center}
  \includegraphics[width=1.0\columnwidth,clip]{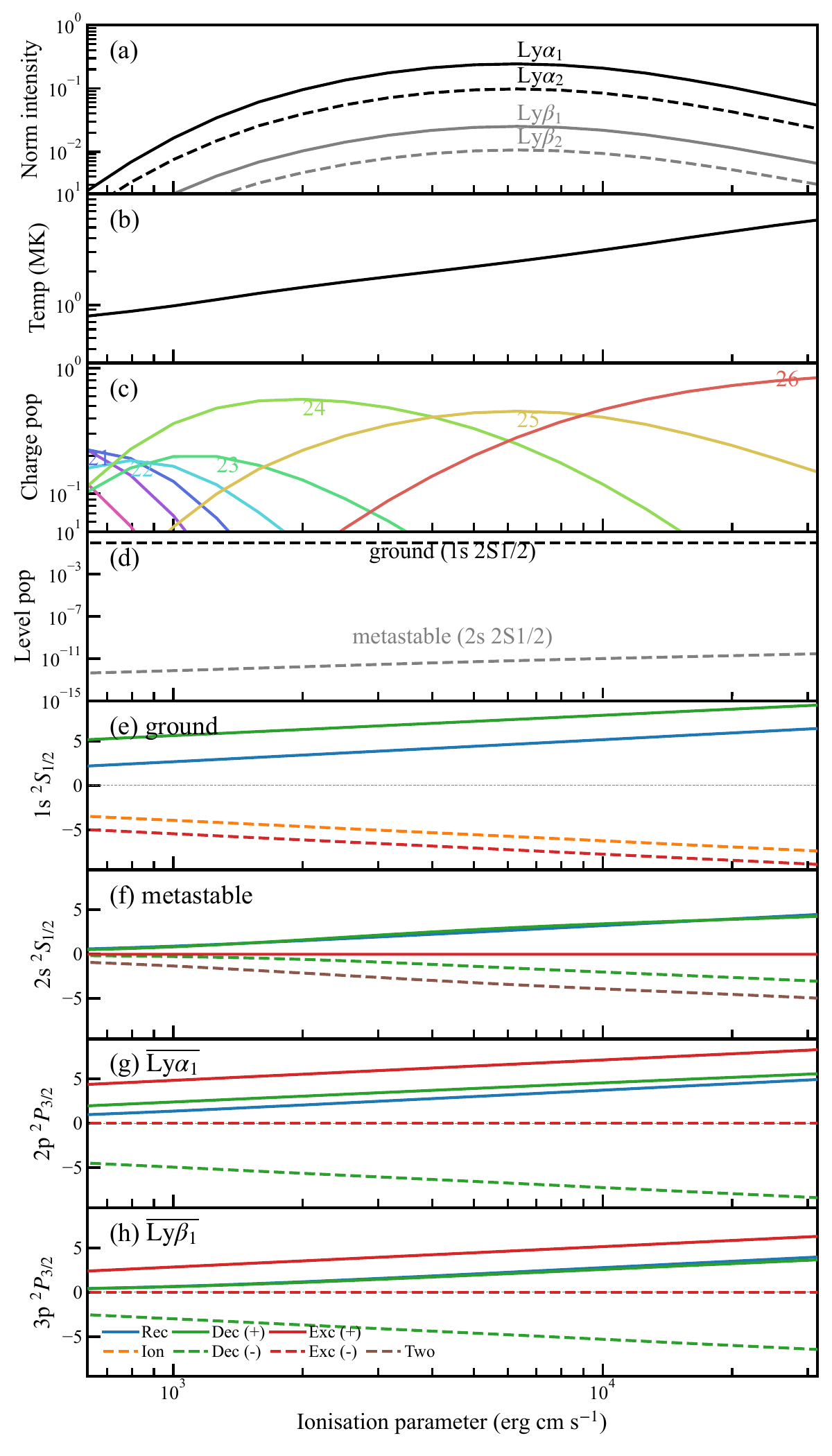}
 \end{center}
 \caption{(a) Fe Ly$\alpha$ and Ly$\beta$ line intensity, (b) electron temperature, (c)
 charge and (d) level populations, and (e)--(h) processes to populate ($y>0$, solid) 
 or de-populate ($y<0$, dashed) each level as a
 function of the ionisation parameter ($\xi$) calculated using \texttt{SPEX}
 \citep{K2024a} at the low-density (10$^{10}$ cm$^{-3}$) and optically-thin limit. In
 (c), the number denotes the number of stripped electrons. In (e)--(h), the rates ($r$)
 are converted with $y=\operatorname{arcsinh}{(r)}$, so that $r$ is shown
 logarithmically as $y \sim \log{|r|}$ at $|r| \gg 1$ while keeping its sign (positive
 for gains and negative for losses). The description of each process can be found in the
 main text.}
 \label{f02}
\end{figure}

We briefly review the atomic levels and processes responsible for the production of
Lyman-series lines. We take Fe as an example, but the physics is the same for other
H-like ions including H and He detailed in \citet{O2006}. The levels involved, as shown
in Fig.~\ref{f01}, are the following:
\begin{enumerate}[label=(\roman*)]
 \item The ground level at $n=1$ (1s $^{2}S_{1/2}$) of H-like ions (Fe$^{+25}$).
 \item The excited levels at $n>1$ ($n$p $^{2}P_{3/2}$ and $n$p $^{2}P_{1/2}$), 
       which are connected to the ground level via an electric dipole (E1) transition.
 \item The metastable level at $n=2$ (2s $^{2}S_{1/2}$), which is the lowest
       level that is not connected to the ground level via an E1 transition.
 \item The ground state of the fully-ionised ions (Fe$^{+26}$) denoted as $n=\infty$. 
\end{enumerate}
Hereafter, we use the designation to indicate the upper or lower level of the 
line $x$ as $\overline{x}$ or $\underline{x}$, respectively; for example, 
$\overline{\mathrm{Ly}\alpha_1}$ for 2p $^{2}P_{3/2}$.

Fig.~\ref{f02} shows the (a) Fe Ly$\alpha$ and Ly$\beta$ line intensity, (b) electron
temperature, (c) charge and (d) level populations, and (e)--(h) the processes to
populate or de-populate each level at the low-density and optically-thin limit of the
photo-ionised plasmas as a function of $\xi$.  The Lyman lines are produced in the range
$\xi = 10^{3}-10^{5}$ and peak at $\xi \sim 10^{3.5}$~erg~cm~s$^{-1}$ (panel a), at
which the Fe$^{+25}$ population peaks (panel c). The electron temperature, at which
radiative heating and cooling balance, is $\approx$1~MK. At such a low temperature, the
collisional processes are sub-dominant. Throughout the range, the ground level is almost
unity and the metastable level population, as well as others, are almost negligible
(panel d).

Different processes dominate to populate each level. For the ground and metastable
levels (panels e and f), the dominant processes to populate each level are the radiative
recombination (RR) directly (`Rec') from Fe$^{+26}$ ($\infty \rightarrow n$) and
the radiative decay from upper levels (`Dec~($+$)').
For the excited levels (panels g and h), these processes contribute as
well, but the radiative excitation from lower levels (`Exc~($+$)') dominates the population. 

For the processes to de-populate each level, the ground level is de-populated by radiative 
excitation to upper levels (`Exc~($-$)') and photo-ionisation to Fe$^{+26}$ (`Ion').
The metastable level is de-populated by two-photon decay (`Two') or the magnetic dipole (M1) 
transition (`Dec~($-$)' in panel f). For the excited levels, 
the radiative excitation to upper levels (`Exc~($-$)') is almost negligible, and 
the radiative E1 transition to the ground state (`Dec~($-$)') is dominant. 
For Ly$\alpha$, the radiative decay ($n=2 \rightarrow 1$) is almost balanced by the radiative excitation 
($1 \rightarrow 2$) in the reverse transitions, making the net value almost negligible. However, this balance breaks for
higher Lyman-series lines. After absorbing a higher Lyman ($1 \rightarrow k$; $k>2$)
photon, an ion does not always re-emit a single photon of the reverse transition ($k
\rightarrow 1$). In some fraction, it instead emits multiple lower-energy photons, for
example a Balmer-series photon ($k \rightarrow 2$) followed by a Ly$\alpha$ photon ($2
\rightarrow 1$), or two-photon decay through the metastable level. Hereafter, we refer
to this conversion of one higher Lyman photon into lower-energy photons as photon
degradation.

These excitation/de-excitation and ionisation/recombination processes balance with each other 
to determine the charge and level populations (panels c and d).

\subsection{Formalism}\label{s2-3}
Based on these physical observations, we implement the Lyman-series lines to \texttt{SKIRT} as
follows: For each element, we sort the relevant levels by energy and assign an index $i$
to each level; for example, $i=0$ for the ground state. We only consider levels up to
$n=10$, beyond which the contribution is insignificant. These levels are discrete and do
not blend with each other. For example, the Fe Ly$\alpha_{1}$ and Ly$\alpha_{2}$ lines
respectively at 6.973 and 6.952~keV have an energy separation (21~eV) much larger than
the typical natural and thermal broadening of $\sim$0.2~eV.

We solve the rate equation for the density $N_{i}$ of the level $i \in \{1,2,\ldots,M\}$
at a steady-state condition by
\begin{align}
 \label{e01}
 \dot{\bm{N}} =\bm{\Gamma}\bm{N} + \bm{S} =0,
\end{align}
where $\bm{N} \equiv (N_1, N_2, \ldots, N_{M})^{\top}$, $\bm{\Gamma}$ is the 
cascade matrix with $\Gamma_{ij}$ representing the contribution from the
level $j$ to $i$, and $\bm{S} \equiv (S_1, S_2, \ldots, S_{M})^{\top}$ is the source
term with $S_{i}$ representing the populating
rates of the level $i$ from the ground ($n=1$) or ionised ($n=\infty$) level. Here, we
exclude the ground level population from $\bm{N}$. Instead, we treat the
transition from the ground level in the source term. The transition from the ionised
level is also treated as the source term.

The cascade matrix is modelled by 
\begin{align}
 \label{e02}
 \Gamma_{ij}N_{j} = \sum_{j=i+1}^{M} A_{ji}N_{j} - N_i \sum_{k=0}^{i-1}A_{ik},
\end{align}
where $A_{ji}$ is the Einstein A coefficient from the level $j$ to $i$. The first term is the 
gain for the level $i$ through the radiative decay from all the upper levels ($j>i$)
proportional to $N_j$ (`Dec~($+$)' in Fig.~\ref{f02}). The second term is the loss from 
the level $i$ through the radiative decay to all the lower levels ($k<i$) proportional 
to $N_i$ (`Dec~($-$)' in Fig.~\ref{f02}). The loss 
through the two-photon decay, which is significant for the metastable
level (`Two' in Fig.~\ref{f02}), is also included in the second term.
We only consider the spontaneous emission in the cascade matrix, which is
dominant over the stimulated emission. By taking the ratio of the two;
\begin{align}
 \label{e03a}
&\frac{B_{ik} J_\nu}{A_{ik}}=\frac{c^2J_\nu}{2 h \nu^3}
\sim\frac{c^2}{2 h \nu^3}\frac{L_\nu}{16 \pi^2 r^2 \Delta \nu}\\
&\sim10^{-7}
\left(\frac{h\nu}{1\,\mathrm{keV}}\right)^{-3}
\left(\frac{n_\text{H}}{10^{12}\,\mathrm{cm^{-3}}}\right)
\left(\frac{\xi}{10^{3}\,\mathrm{erg\,cm\,s^{-1}}}\right)
\left(\frac{h\Delta\nu}{1\,\mathrm{eV}}\right)^{-1}, \nonumber
\end{align}
where $B_{ik}$ is the Einstein B coefficient from the level $i$ to $k$, $J_\nu$ and
$L_\nu$ are the radiation field strength and the luminosity at the line transition
frequency ($\nu$), and $\Delta \nu$ is the line width. Here, we used the Einstein relation
$B_{ik}/A_{ik} = c^2/2h\nu^3$ and substituted $L_\nu$ with $L_\mathrm{X}$ to relate the ratio
with the ionisation parameter ($\xi$). This estimate shows that $B_{ik}J_\nu/A_{ik}\ll1$
is satisfied in the plasmas of our interest, justifying our assumption to neglect the
stimulated transitions.

The source term is modelled by 
\begin{align}
 \label{e03}
 S_i = n_{\mathrm{e}} n_{\infty} \alpha^\mathrm{RR}_i(T) + n_0 B_{0i} J(\nu=\nu_{i0}).
\end{align}
The first term is for the recombination from the fully-ionised ions (`Rec' in Fig.~\ref{f02}), in which
$n_{\mathrm{e}}$, $n_0$ and $n_{{\infty}}$ are the densities of the electrons, ground level of 
H-like ions, fully-ionised ions, respectively, and $\alpha^{\mathrm{RR}}_i(T)$ is the partial radiative
recombination coefficient to the level $i$ as a function of the electron temperature
($T$). The second term is for the radiative excitation from the ground to the level $i$
of the H-like ions (`Exc($+$)' in Fig.~\ref{f02}), which is proportional to the radiation field $J$ at the
corresponding frequency ($\nu_{i0}$) with the Einstein $B$ coefficient of the transition.

We treat two channels separately. One is the
resonance scattering (RS) and the other is the radiative recombination (RR). Because of
the linearity, Eq.~(\ref{e01}) can be divided into 
\begin{align}
 \bm{\Gamma}\bm{N}_{\mathrm{RS}} + \bm{S}^{\mathrm{RS}} &= \bm{0},\label{e04} \\
 \bm{\Gamma}\bm{N}_{\mathrm{RR}} + \bm{S}^{\mathrm{RR}} &= \bm{0},\label{e05}
\end{align}
where $\bm{N}_{\mathrm{RS}}$ and $\bm{N}_{\mathrm{RR}}$ are the separated populations of
the two channels. The source term is also separated into $S^{\mathrm{RS}}_i \equiv n_0
B_{0i} J(\nu=\nu_{i0})$ and $S^{\mathrm{RR}}_i \equiv n_{\mathrm{e}} n_{\infty}
\alpha^\mathrm{RR}_i(T)$ in Eq.~(\ref{e03}). The solutions of these equations satisfy
Eq.~(\ref{e01}) by substituting $\bm{N} = \bm{N}_{\mathrm{RS}} + \bm{N}_{\mathrm{RR}}$.

This separation is physically motivated as the two channels contribute quite differently
for the Ly$\alpha$ photons. In RS, the absorption and re-emission of Ly$\alpha$ photons
repeat numerous times with a large oscillator strength. A Ly$\alpha$ photon behaves as
if it were scattered by H-like ions, hence the process is called resonance
scattering. The scattering does not contribute to the net production of Ly$\alpha$
photons, but is important to determine the line profile. Ly$\alpha$ photons diffuse not only 
in space but also in frequency through numerous consecutive scattering events. When
a photon is scattered into a frequency far from the line centre, the optical depth drops
significantly, and the line photon can escape the medium \citep{H1980}. The emergent line
profile is strongly affected by this process \citep{H1973,N1991,D2006}. Also, resonance 
scattering induces linear polarisation \citep{H1947}.

The processes for the higher Lyman photons ($n>1$) are similar, but with the notable
difference of photon degradation. For every degraded higher Lyman photon, a lower Lyman
photon, including Ly$\alpha$, is produced.

In RR, the H-like ions are ionised by a photon exceeding the ionisation threshold
(9.277~keV in case of Fe$^{+25}$), which recombines with a free electron back to
Fe$^{+25}$. When the recombination takes place to an excited level and decays through
the $\overline{\mathrm{Ly}\alpha}$ level, it contributes to the production of the
Ly$\alpha$ photons.

\section{Implementation}\label{s3}
\texttt{SKIRT} tracks individual packets of photons until they escape from the system or
are absorbed (i.e., destructed). Because we are interested in the Lyman-series photons,
we make a simplification when a higher Lyman photon is degraded. The degradation
produces a lower Lyman photon and one or more photons of lower energies. We only keep
track of the Lyman photon and neglect the others. As shown in Eq.~(\ref{e03a}), induced
radiative transitions among excited levels can be neglected within the scope of this
work. These neglected lower-energy photons are, therefore, not expected to
significantly affect the level populations and the resultant Lyman-series line profiles.

We implement the prominent E1 transition lines between the $n$p~$^2P_{3/2,1/2}$ and 
the ground level (1s $^2S_{1/2}$) for $n=2$--$10$, i.e., 18 Lyman-series lines for each 
element with $Z=1$--$30$ (Ly$\alpha_1$, Ly$\alpha_2$, Ly$\beta_1$, Ly$\beta_2$, $\ldots$, 
Ly$\theta_1$, Ly$\theta_2$, Ly$\iota_1$, Ly$\iota_2$). We also implement the M1 transition
lines between the metastable level (2s $^2S_{1/2}$) and the ground level for $Z\geq14$.
These lines are much weaker than the Ly$\alpha$ lines, but are comparable to the higher E1 
lines and thus should not be neglected.

We solve the level populations for the H-like ions ($N_i$; $i>0$) using the rate
equations (Eqs.~(\ref{e04}) and (\ref{e05})), but we do not solve the charge populations. These
can be calculated externally by other RT solvers, such as \texttt{SPEX} \citep{K2024a},
\texttt{Cloudy} \citep{G2025b}, or \texttt{XSTAR} \citep{K1994}, as in Fig.~\ref{f02}
(c). These two-stream RT codes are commonly used in X-ray studies to solve the radiative
heating and cooling balance for the electron temperature ($T$), ionisation and
recombination balance for the charge population, and excitation and de-excitation
balance for the level population under a given $\xi$ and SED of the photo-ionising
source. We use their result to set $n_0$, $n_{{\infty}}$, and $T$ 
in each radial cell of the spherically symmetric \texttt{SKIRT} model.
Since 1D RT codes solve the photo-ionisation equilibrium, this setup realises a
balance between the photo-ionisation rate and recombination rate under 
the local radiation field in each cell, which therefore justifies the scheme that a single
photo-ionisation event is followed by a single recombination event.

The separation into the two channels is also beneficial as we can exploit the existing
\texttt{SKIRT} implementations. For RS, the implementation of the frequency
redistribution is described by \citet{C2021}, while that of the M\"{u}ller matrix for
the polarisation angular redistribution is by \citet{P2017}. For RR, the process is
implemented in the same way as the inner-shell ionisation of cold matter followed by
fluorescence as described by \citet{M2023b,M2024}.

\citet{C2021} note that the \texttt{SKIRT} results for RS of the H Ly$\alpha$ line at
very high optical depths depend on the type and resolution of the spatial grid. This is
not an issue for the present applications, as the optical depth is typically
significantly smaller for metallic lines. The core-skipping mechanism implemented in
\texttt{SKIRT} for acceleration purposes of RS has been turned off throughout this work
as it is incompatible with the presented implementation in general.

Below, we discuss the amplitude, branching ratio, and
redistribution for each of the two channels separately. 
We retrieved the atomic data from \texttt{SPEX} v.3.08.01 \citep{K2024a}
for all levels below $n=10$ for all elements of $Z=$1--30.

\subsection{Resonance scattering}\label{s3-1}
We consider resonance scattering photons between the ground level and the excited level $i$
with an intensity $\bm{I}_i (\nu, \hat{\bm{k}}) = (I_i (\nu, \hat{\bm{k}}), Q_i (\nu,
\hat{\bm{k}}), U_i (\nu, \hat{\bm{k}}))^{\top}$ using the Stokes parameters, where
$\nu$ is the frequency and $\hat{\bm{k}}$ is the unit vector for the propagating
direction. This is scattered into $\bm{I}_i (\nu^{\prime}, \hat{\bm{k}}^{\prime})$ with
a different primed frequency and direction, which is described by the redistribution matrix
$\bm{R}(\nu, \hat{\bm{k}}, \nu^{\prime}, \hat{\bm{k}}^{\prime})$ in the $3 \times 3$
format for the three Stokes parameters.

It is commonly the case that the redistribution over the frequency ($\nu \rightarrow
\nu^{\prime}$) happens within the profile of the same line. However, because we also
consider the photon degradation, we generalise the matrix as $\bm{R}_{ij}$ to allow
redistribution across lines of different upper levels ($j$ to $i$). This yields the
radiative transfer equation as
\begin{equation}
 \label{e06}
  \frac{\mathrm d \bm{I}_i(\nu', \hat{\bm{k}}')}{\mathrm ds}=
  \sum_{j \ge i}\bm{R}_{ij}(\nu, \hat{\bm{k}}, \nu', \hat{\bm{k}}')
  \, P_{ij}\,\alpha^\text{RS}_j(\nu, T)\,\bm{I}_j(\nu, \hat{\bm{k}}) ,
\end{equation}
where $\alpha^\text{RS}_j(\nu, T)$ is the amplitude, $P_{ij}$ is the branching ratio to
represent the degradation probability from $j$ to $i$, and $\bm{R}_{ij}(\nu, \hat{\bm{k}},
\nu', \hat{\bm{k}}')$ is the generalised redistribution matrix.

\subsubsection{Amplitude}\label{s3-1-1}
The amplitude of the RS is given by the bound--bound absorption cross-section
\citep{R1979} as
\begin{equation}
 \label{e07}
  \alpha^\text{RS}_{i}(\nu, T) = n_{0} \left(f_{i}\frac{\pi e^{2}}{m_{\mathrm{e}}c}\right) \phi(\nu-\nu_{i}, T),
\end{equation}
where $n_0$ is the density of the ground level, $f_{i}$ is the oscillator
strength from the ground to the level $i$, $\nu_{i}$ is the line centre frequency, and
$\phi(\nu, T)$ is the line profile normalised to unity when integrated over $\nu$, and others follow
the convention. We consider both the natural and thermal broadening with a temperature
$T$, thus $\phi(\nu, T)$ is a Voigt profile that also depends on $T$.

\subsubsection{Branching ratio}\label{s3-1-2}
When a H-like ion in the ground state is radiatively excited to the level $j$, the
subsequent radiative cascades determine the equilibrium level population $N_i$, where
$i<j$. The population of each level is obtained by solving Eq.~(\ref{e04}) in which the source term
includes only the radiative excitation $0 \rightarrow j$ as
\begin{align}
      S^{\mathrm{RS}}_k = 
      \begin{cases}
            n_0 B_{0j} J(\nu=\nu_{j0}) & (k=j), \\
            0 & (k \neq j).
      \end{cases}
\end{align}
The degradation probability from $j$ to $i$ is then given as the
fraction of the $i \rightarrow 0$ transition rate over the $0 \rightarrow j$
radiative excitation rate by
\begin{align}
 P_{ij}=\frac{N_i A_{i0}}{n_0 B_{0j} J(\nu=\nu_{j0})}.
\end{align}

\subsubsection{Redistribution}\label{s3-1-3}
We decompose the redistribution matrix into two parts for the redistribution
over the frequency and direction.
\begin{equation}
 \label{e09}
  \bm{R}_{ij}(\nu, \hat{\bm{k}}, \nu^{\prime}, \hat{\bm{k}}^{\prime}) = f_{ij}(\nu,
  \nu^{\prime}) \bm{g}_{ij}( \hat{\bm{k}}, \hat{\bm{k}}^{\prime})
\end{equation}
For the frequency redistribution $f_{ij}(\nu, \nu^{\prime})$, we follow the scheme by
\citet{H1962}. For RS without photon degradation ($i=j$), the absorbed and re-emitted
photons are coherent. Hence the partial redistribution (type II) takes place, in which 
the photon frequency is redistributed around the absorbed photon frequency with thermal 
broadening. The energy transfer through recoil is neglected, as in \cite{C2021}, because it 
is negligible compared with the Doppler broadening at the temperatures considered in this work.
For RS with photon degradation ($j>i$), the coherency does not hold because
the absorbed photon is re-emitted through a different transition.
We therefore adopt the complete redistribution, in which the photon frequency
is redistributed around the line centre frequency.

For the direction redistribution $\bm{g}_{ij}( \hat{\bm{k}}, \hat{\bm{k}}^{\prime})$, we
follow the scheme by \citet{H1947,C1960}. Because the redistribution function is
different for the three Stokes parameters, $\bm{g}_{ij}(\hat{\bm{k}},
\hat{\bm{k}}^{\prime})$ is a $3 \times 3$ matrix. In the global frame of the simulation,
the matrix is written as
\begin{equation}
\bm{g}_{ij}(\hat{\bm{k}},\hat{\bm{k}}')=\bm{M}(\theta)\,\bm{L}(\varphi),
\end{equation}
where $\bm{L}(\varphi)$ represents the rotation matrix from the global frame to the
local frame of each scattering event defined by $\hat{\bm{k}}$ and $\hat{\bm{k}}'$,
$\bm{M}(\theta)$ is the M\"{u}ller matrix, and $\theta$ is the angle between
$\hat{\bm{k}}$ and $\hat{\bm{k}}'$ \citep{P2017}. The matrix is given as a linear
combination of monopole and dipole distribution.
\begin{eqnarray}
 \label{e10}
  \bm{M}(\theta) &=& E_1 \bm{M}^{(m)} (\theta) +E_2 \bm{M}^{(d)} (\theta) \nonumber \\
              &=& E_1 \frac{1}{2}\left(\begin{matrix} 1 & 0 & 0 \\ 0 & 0 & 0 \\ 0 & 0 & 0 \end{matrix} \right)\\
            &+& E_2 \frac{3}{8}\left(\begin{matrix} 1+\cos{\theta}^{2} &
				     -\sin{\theta}^{2} & 0 \\ -\sin{\theta}^{2} &
				     1+\cos{\theta}^{2} & 0 \\ 0 & 0 & 2\cos{\theta}
				     \end{matrix} \right)\nonumber
\end{eqnarray}
Here, $\bm{M}^{(m)}$ and $\bm{M}^{(d)}$ are the monopole and dipole terms normalised to satisfy
$\int_{0}^{\pi} M_{11}(\theta)\sin{\theta}d\theta = 1$ in each, and $E_1$
and $E_2$ ($E_1 + E_2=1$) are their relative weights. For RS without photon degradation
($i=j$), the value depends on the quantum number $J$ of the lower level and its
difference $\Delta J$ with the upper level \citep{H1947}. For example, Ly$\alpha_1$
($J=1/2$ and $\Delta J=1$) has $E_1=0.5$ and $E_2=0.5$ and Ly$\alpha_2$ ($J=1/2$ and
$\Delta J=0$) has $E_1=1$ and $E_2=0$. For RS with photon degradation ($j>i$), we
assumed the monopole distribution only ($E_1=1$ and $E_2=0$).

\subsection{Radiative recombination}\label{s3-2}
We consider the ionising photon $\bm{I}(\nu, \hat{\bm{k}})$ at $\nu > \nu_{\infty}$,
where $\nu_{\infty}$ is the ionisation threshold of H-like ions. When H-like ions are
ionised to fully-ionised ions, they are assumed to recombine immediately with a free
electron to go back to the H-like ion at the ground level or at an excited level, which
eventually cascades down to the ground level. We focus only on the Lyman-series photons
as a product of the cascade from the level $i$ to the ground level. A single ionising
photon yields a single Lyman-series photon, thus this channel can be described as an
inelastic scattering.  The radiative transfer equation reads
\begin{equation}
 \label{e11}
 \frac{\mathrm d \bm{I}_i(\nu^{\prime}, \hat{\bm{k}}^{\prime})}{\mathrm{d}s} = 
 \bm{R}_i(\nu, \hat{\bm{k}}, \nu^{\prime}, \hat{\bm{k}}^{\prime})\,
 P_i(T)\,\alpha^\text{PI}(\nu)\,\bm{I}(\nu, \hat{\bm{k}}),
\end{equation}
where $\bm{I}_{i}$ is the resulting Lyman photon intensity with an upper level $i$,
$\alpha^{\mathrm{PI}}(\nu)$ is the photo-ionisation amplitude, $P_{i}(T)$ is the probability that
a single photo-ionisation event yields a Lyman $i \rightarrow 0$ photon, and $\bm{R}_{i}$
is the redistribution matrix over the frequency and direction in the form of a $3 \times
3$ matrix for the three Stokes parameters.

\subsubsection{Amplitude}\label{s3-2-1}
The amplitude is the product of the ground state density $n_0$ and the parameterised 
K-shell photo-ionisation cross section \citep{V1995},
\begin{align}\label{e12}
  \alpha^\text{PI}(\nu) &= n_{0}\sigma_0 F(\nu),
\end{align}
where
\begin{align}\label{e13}
 F(\nu) &= \left[(\nu/\nu_0-1)^2 + y_w^2 \right] \, (\nu/\nu_0)^{0.5P - 5.5} \left( 1 + \sqrt{(\nu/\nu_0)/y_a} \right)^{-P}.
\end{align}
Here, $\sigma_0, \nu_0, y_w, y_a$ and $P$ are the fitting parameters
for the ground state of all ions with $Z=$1--30, the values of which are available online
\footnote{\url{https://www.pa.uky.edu/~verner/photo.html}.}.

\subsubsection{Branching ratio}\label{s3-2-2}
When a fully-ionised ion at $i=\infty$ recombines with an electron, it leaves a H-like
ion at an excited ($i>0$) or the ground ($i=0$) level; the former case is followed by 
subsequent radiative cascades. The equilibrium level population $N_i$ is then obtained by solving
Eq.~(\ref{e05}). The level-resolved partial recombination rate $\alpha^{\mathrm{RR}}_i(T)$ 
averaged over the Maxwellian distribution
of electrons with a temperature $T$ is given parametrically by \cite{M2016a} based on
the Atomic Data and Analysis Structure (ADAS) products \citep{B2006}, which we utilise. For fully-ionised ions, the
dielectronic recombination does not occur. The branching ratio is then given as the
fraction of the $i \rightarrow 0$ transition rate over the total recombination rate by
\begin{align}
 \label{e14}
 P_i(T)=\frac{N_i A_{i0}}{n_{\mathrm{e}} n_{\infty} \sum_k \alpha^{\mathrm{RR}}_k(T)}.
\end{align}

\begin{figure}
 \begin{center}
  \includegraphics[width=1.0\columnwidth,clip]{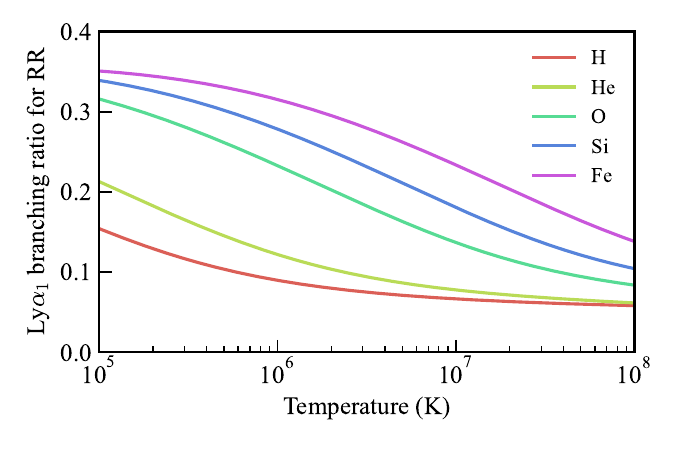}
 \end{center}
 \caption{Branching ratio that a photo-ionising photon produces a single
 Ly$\alpha_1$ photon after radiative recombination and cascade for several selected
 elements.}
 \label{f03}
\end{figure}

The computed branching ratio for Ly$\alpha_1$ is shown in Fig.~\ref{f03} for several
elements. The probability decreases for higher temperatures because the recombination
directly to the ground level increases, which does not contribute to the production of
Lyman-series photons.

\subsubsection{Redistribution}
The redistribution function is decomposed into the redistribution over the frequency and
direction in the same way as the RS channel (Sect.~\ref{s3-1-3}),
\begin{equation}
 \label{e15}
  \bm{R}_{i}(\nu, \hat{\bm{k}}, \nu^{\prime}, \hat{\bm{k}}^{\prime}) = f_{i}(\nu,
  \nu^{\prime}) \bm{g}_{i}( \hat{\bm{k}}, \hat{\bm{k}}^{\prime}).
\end{equation}
The incident ionising photon and the resultant Lyman photon are not coherent. Therefore,
for the frequency redistribution $f_{i}(\nu, \nu^{\prime})$, we assume that the
resultant photon is distributed around the line centre by thermal broadening. For the
direction redistribution $\bm{g}_{i}( \hat{\bm{k}}, \hat{\bm{k}}^{\prime})$, we adopt
the monopole distribution.

\section{Verification}\label{s4}
We now verify the implementation in Sect.~\ref{s3} by comparing synthesised spectra using
\texttt{SKIRT} with analytical solutions or results of a different RT code. As we relied
on \texttt{SPEX} in the implementation, we use \texttt{Cloudy} c25 release
\citep{G2025b}, based on a two-stream solver, for comparison. In the former half
(Sect.~\ref{s4-1}), we examine the microphysics separately for the amplitude
(Sect.~\ref{s4-1-1}), branching ratio (Sect.~\ref{s4-1-2}), and redistributions
(Sect.~\ref{s4-1-3}) with a dedicated setup for each to verify that the code is implemented
as intended. In the latter half (Sect.~\ref{s4-2}), we investigate three representative RT
effects that are actually observable with XRISM --- distortion of a line profile
(Sect.~\ref{s4-2-1}), the Lyman decrement (Sect.~\ref{s4-2-2}), and formation of P Cygni
profile (Sect.~\ref{s4-2-3}). The simulation setups are summarised in Table~\ref{t01}.
We use $10^{10}$ photon packets and 100 radial grid cells in each simulation. 

\begin{table*}
 \centering
 \caption{Setups of the \texttt{SKIRT} simulations in Sect.~\ref{s4}.}
 \label{t01}
 \begin{tabular}{cll}
  \hline
  Subsections & Shell parameters & Incident source  \\
  \hline
  Sect.~\ref{s4-1-1}&$N_\mathrm{Fe^{+25}}=10^{16,18}~\mathrm{cm^{-2}}, T=10^6~\mathrm{K}$ & 
	  central, flat SED, unpolarised\\
  Sect.~\ref{s4-1-2}&$n_\mathrm{H}=10^{10}~\mathrm{cm^{-3}}, N_\mathrm{H}
      =10^{20}~\mathrm{cm^{-2}}, \xi=10^{3.5}~\mathrm{erg~s^{-1}~cm}$&central, flat SED
	  with truncation, unpolarised\\
  Sect.~\ref{s4-1-3}&$\tau_0=0.01, T=10^6~\mathrm{K}$&(a) central, monochromatic, unpolarised\\
                 &                                &(b) outside the shell, monochromatic, polarised\\
  Sect.~\ref{s4-2-1}&$\tau_0=0.1,10,10^3, T=10^6~\mathrm{K}$&central, monochromatic, unpolarised \\
  Sect.~\ref{s4-2-2}&$n_\mathrm{H}=10^{10}~\mathrm{cm^{-3}}, N_\mathrm{H}
      =10^{20-24}~\mathrm{cm^{-2}}, \xi=10^{3.5}~\mathrm{erg~s^{-1}~cm}$&central, flat SED, unpolarised \\
  Sect.~\ref{s4-2-3}&$\tau_0=0.1, T=10^6~\mathrm{K}$&central, flat SED, unpolarised \\
  \hline
 \end{tabular}
 \tablefoot{The medium is a shell consisting of Fe$^{+25}$ in all simulations, and $\tau_0$ denotes
 the optical depth of Fe Ly$\alpha_1$. Further details are given in the main text.}
\end{table*}

\subsection{Microphysics}\label{s4-1}
\subsubsection{Amplitude}\label{s4-1-1}
\begin{figure}
 \begin{center}
  \includegraphics[width=1.0\columnwidth,clip]{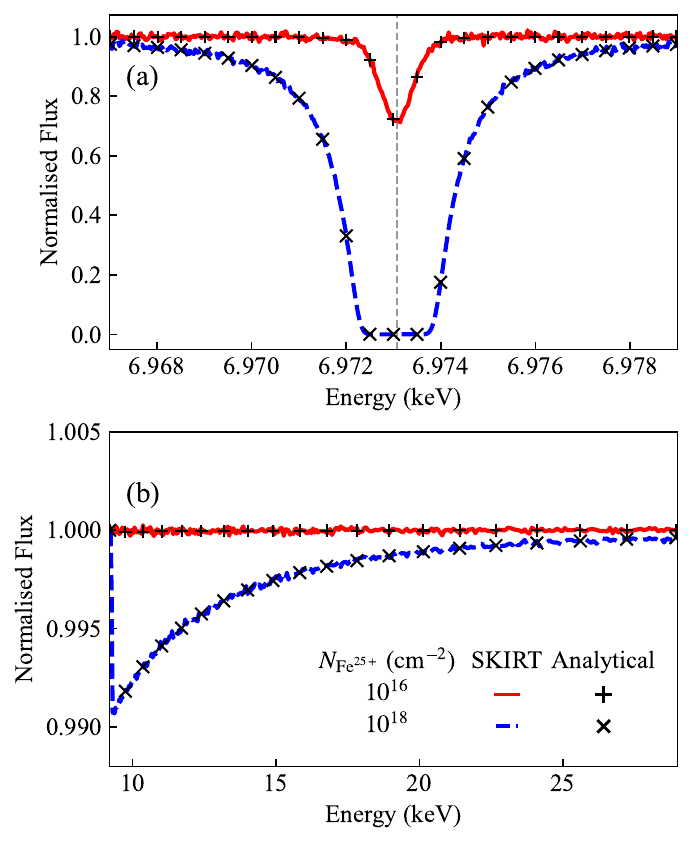}
 \end{center}
 \caption{(a) Fe Ly$\alpha_1$ absorption (RS cross-section) and (b) Fe$^{+25}$
 photo-ionisation absorption edge (RR cross-section). The transmitted component
 calculated with \texttt{SKIRT} for $N_\mathrm{Fe^{+25}}=10^{16}$ (red solid) and
 $10^{18}~\mathrm{cm^{-2}}$ (blue dashed) are shown with the corresponding analytical
 profiles (black plus and cross, respectively). The grey vertical line in (a) represents
 the rest-frame energy of the line.}
 \label{f04}
\end{figure}

The amplitudes of the RS (Eq.~(\ref{e07})) and the RR (Eq.~(\ref{e12})) are given in the form
of a cross-section, which appears in the transmitted component of synthesised
spectra. We ran a \texttt{SKIRT} simulation for a uniform gas sphere consisting
only of Fe$^{+25}$ at a 10$^{6}$~K temperature. The incident source is point-like at the
centre of the sphere with isotropic and unpolarised emission with a flat SED. Two
runs were made for different column densities of Fe$^{+25}$: $10^{16}$ and
$10^{18}~\mathrm{cm^{-2}}$, for which the Ly$\alpha$ line is optically thin and thick,
respectively. The result is shown in Fig.~\ref{f04} (a) for the line absorption and (b)
photo-ionisation absorption edge. Both results match well with the analytical results.

\subsubsection{Branching ratio}\label{s4-1-2}
The branching ratio can be best examined using the Lyman-series lines in the scattered
component of the synthesised spectra, as their intensity is proportional to their upper
level populations, which depends on the branching ratio of both RS and RR channels.

We first performed a \texttt{Cloudy} simulation of a shell with a uniform density
$10^{10}~\mathrm{cm}^{-3}$ and a column density $10^{20}~\mathrm{cm}^{-2}$, which is
photo-ionised by the incident source placed at the centre with $\xi =
10^{3.5}$~erg~s$^{-1}$~cm. The incident source is point-like and emits isotropic and
unpolarised radiation with a flat SED. For individual verification of the two channels,
we truncated the SED at the Fe$^{+25}$ K edge energy, below which RS is in effect and
above which RR is in effect.

\begin{figure}
 \begin{center}
  \includegraphics[width=1.0\columnwidth,clip]{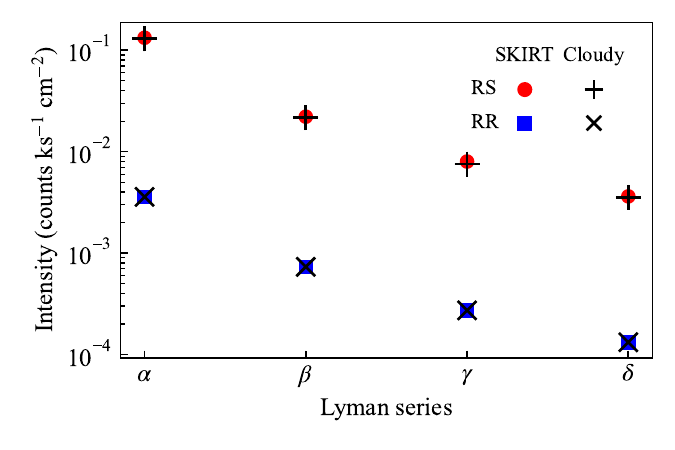}
 \end{center}
 \caption{Photon flux of the Fe Lyman-series lines from the RS and RR channels calculated
 with \texttt{SKIRT} (red circle and blue square, respectively) and \texttt{Cloudy}
 (black plus and cross, respectively). The incident source has
 $L_{\mathrm{X}}=10^{38}$~erg~s$^{-1}$ and the lines are observed at a 2.5~kpc
 distance.}
 \label{f05}
\end{figure}

We derived the spatial distribution of the temperature and the ion density with
\texttt{Cloudy}, which we used for the \texttt{SKIRT} runs. We compare the absolute
intensity of the Lyman-series lines (the fine-structure doublet is summed) between
\texttt{Cloudy} and \texttt{SKIRT} in Fig.~\ref{f05}, which matches very well for both
channels.

\subsubsection{Redistribution}\label{s4-1-3}
\begin{figure}
 \begin{center}
  \includegraphics[width=1.0\columnwidth,clip]{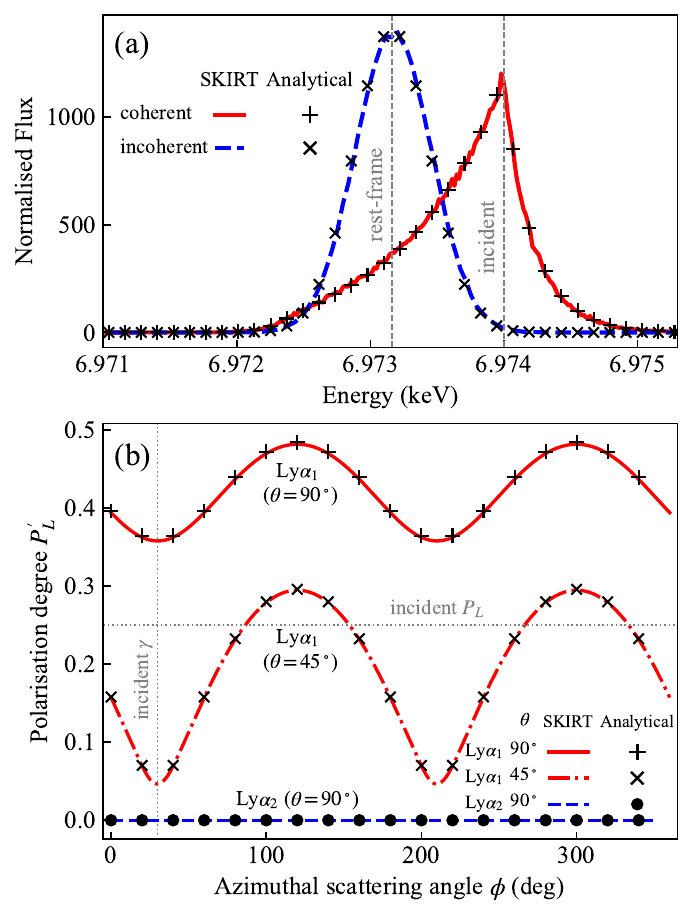}
 \end{center}
 \caption{(a) Frequency redistribution for coherent (red solid curve) and incoherent
 (blue dashed curve) scattering of Fe Ly$\alpha_1$ calculated with
 \texttt{SKIRT}. Analytical solutions are compared in black: the type II redistribution
 function of \cite{H1962} (coherent), and the Maxwellian distribution with a $10^6$~K
 temperature (incoherent). The vertical dashed lines mark the rest-frame energy of the
 line and the energy of the incident photons for the coherent scattering.
 (b) Polarisation degree as a function of the azimuthal scattering angle $\phi$ for
 Ly$\alpha_1$ observed at $\theta=90^{\circ}$ and 45$^{\circ}$ and for Ly$\alpha_2$ at
 $\theta=90^{\circ}$ calculated with \texttt{SKIRT}. Analytical solutions
 (Eq.~(\ref{e16})) are compared with points. The incident polarisation degree $P_{L}=0.25$
 and angle $\gamma=30^{\circ}$ are also shown with dotted lines.}
 \label{f06}
\end{figure}

\paragraph{Frequency redistribution}
For the frequency redistribution, we tested the Fe Ly$\alpha_1$ emission profile
separately for the coherent and incoherent scattering. We ran a \texttt{SKIRT} simulation
for a spherical uniform gas consisting only of Fe$^{+25}$ with a 10$^{6}$~K temperature
and a 0.01 optical depth at the line centre. The incident source is point-like at the
centre of the sphere with isotropic, unpolarised, and monochromatic radiation.

We had two different monochromatic energies as an input, so that a process to produce
either the coherent or incoherent scattering is invoked.  For the coherent scattering, the
incident energy was set at 6.974~keV, which is slightly offset from the line
centre. This produces a Ly$\alpha_1$ photon by coherent scattering.  For the incoherent
scattering, the incident energy was set at the Ly$\gamma_1$ line centre
(8.701~keV). This produces a Ly$\alpha_1$ photon through the degradation, which is
incoherent.

The results are shown in Fig.~\ref{f06} (a). We confirmed that coherent scattering
redistributes photons around the incident energy, whereas incoherent scattering
redistributes them around the rest-frame energy of the line centre.  We also show
analytical expressions: the type II redistribution function from \citet{H1962} for
coherent scattering, and the Maxwellian distribution for incoherent scattering. The
\texttt{SKIRT} results match the analytic solutions well in both cases.

\paragraph{Direction redistribution}
For direction redistribution, we tested the line polarisation degree separately for
the Ly$\alpha_{1}$ or Ly$\alpha_{2}$ lines.
We ran \texttt{SKIRT} simulations using the setup described above with several changes:
(i) the incident energy is fixed at the line centre of either Fe Ly$\alpha_{1}$ or Ly$\alpha_{2}$ line,
(ii) the direction of incident photons is set as a laser-like parallel beam from
the bottom of the sphere, and (iii) the incident photons are linearly polarised with a
polarisation degree of $P_L=0.25$ and a polarisation angle of $\gamma=30^{\circ}$. In this
case, the observed polarisation degree is theoretically given by Eq.~(18) of \cite{V2024}.
\begin{align}
 \label{e16}
P'_L &=
\frac{\sqrt{\left( M_{12} + M_{22} P_L \cos 2(\gamma - \phi) \right)^2
+\left( M_{33} P_L \sin 2(\gamma - \phi) \right)^2}}
{M_{11} + M_{12} P_L \cos 2(\gamma - \phi)}.
\end{align}
Here, $M_{ij} \equiv M_{ij}(\theta)$ denotes the $(i,j)$ component of the M\"{u}ller
matrix (Eq.~(\ref{e10})), and $\phi$ is the azimuthal scattering angle. Note that
polarisation is produced ($P'_L>0$) even for $P_{L}=0$ in Eq.~(\ref{e16}). The results are
shown in Fig.~\ref{f06} (b) for Ly$\alpha_1$ seen from $\theta =45^{\circ}$ and
90$^{\circ}$ and Ly$\alpha_2$ seen from $\theta =90^{\circ}$. For Ly$\alpha_2$, no
polarisation was observed as expected, because the scattering is isotropic
(Eq.~(\ref{e10})). For Ly$\alpha_1$, the polarisation degree is
modulated by $2(\gamma-\phi)$ as expected, because the scattering has a dipole
component (Eq.~(\ref{e10})). The polarisation degree is also higher for larger
$\theta$ as the weight of the dipole component increases. All of the \texttt{SKIRT} results agree well with the
analytical expectations.

\subsection{Radiative transfer effects}\label{s4-2}
\subsubsection{Distortion of line profile}\label{s4-2-1}
Line photons of a strong oscillator suffer high optical depth at the line centre
($\tau_0$) in photo-ionised plasmas around compact objects. For example, the Fe
Ly$\alpha$ line reaches $\tau_0=1$ for a plasma column of only $1.7\times
10^{21}$~cm$^{-2}$ \citep{T2025a}, which is much smaller than the columns that are
actually observed \citep{A2025a}. When $\tau_0\gg1$, the line photon can only escape from
the system when it is scattered into a frequency far from the centre, thus the emergent
line profile is severely distorted. However, none of the two-stream RT codes commonly
used in X-rays have the capability to calculate this distortion accurately.  They use
the escape probability scheme, in which the observed line emission is evaluated by
integrating the local line emission weighted by the probability that a photon escapes
the medium.  Therefore, the diffusion over frequencies is not calculated, which is
computationally expensive. The \texttt{SKIRT} MC-RT code is better suited for this
problem.

The emergent profile can be analytically solved for a spherical geometry with a static and
uniform density distribution \citep{D2006}. The asymptotic profile at
$\tau_0 \gg 1$ is given by
\begin{equation}
 \label{e17}
 J(x)=\frac{\sqrt{\pi}}{\sqrt{24}\,a\tau_0}\,
  \frac{x^{2}}{1+\cosh\!\left[\sqrt{\frac{2\pi^{3}}{27}}\,
			 \frac{|x|^{3}}{a\tau_0}\right]},
\end{equation}
where $a \equiv \frac{\Gamma}{4\pi \Delta \nu_{\mathrm{D}}}$ is the Voigt parameter
defined as the ratio between the natural ($\Gamma = \sum_{j<i} A_{ji}$) and thermal
($\Delta \nu_{\mathrm{D}} = \frac{\nu_0}{c}\sqrt{\frac{2kT}{m_Z}}$) broadening for the
temperature $T$ and the ion mass $m_{Z}$ and the dimensionless frequency $x \equiv
(\nu-\nu_0)/\Delta \nu_{\mathrm{D}}$.

\begin{figure}
 \begin{center}
  \includegraphics[width=1.0\columnwidth,clip]{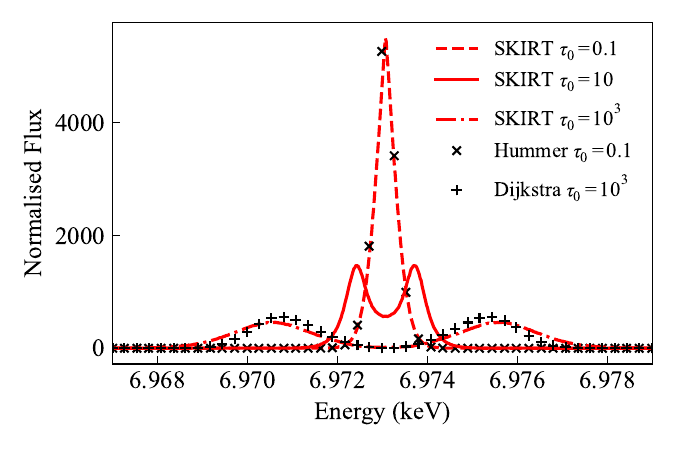}
 \end{center}
 \caption{Ly$\alpha_1$ line profile calculated using \texttt{SKIRT} (red) for
 $\tau_0=0.1$ (dashed), 10 (solid), and 10$^{3}$ (dashed-and-dotted) with
 $T=10^{6}$~K. The analytical expressions for $\tau_0=0.1$ \citep{H1962} and 
 $\tau_0=10^3$ \citep{D2006} are also shown as black crosses and pluses, respectively.}
 \label{f07}
\end{figure}

We ran \texttt{SKIRT} simulations for a sphere consisting only of Fe$^{+25}$ ions of a
uniform density at $T=10^{6}$~K. The optical depth of the Ly$\alpha_1$ line was changed
as $\log{\tau_0} \in \{-1, 1, 3\}$. The resultant line profile is shown in
Fig.~\ref{f07}. For the optically-thin case ($\log{\tau_0}=-1$), the profile follows the
type II redistribution function of \cite{H1962}. However, as $\tau_0$ increases, 
the profile is more and more distorted,
which is reproduced by \texttt{SKIRT}'s calculation of the redistribution over
frequencies at individual RS events. For the optically-thick case ($\log{\tau_0}=3$),
the \texttt{SKIRT} result is close to the asymptotic expression by \citet{D2006}. The
dispersion in the red and blue wing of the distorted line profile amounts to $\sim$6~eV
for the Fe Ly$\alpha$ line, which can be easily recognised with contemporary
high-resolution X-ray spectrometers, but is often inappropriately represented as a
single Voigt profile broadened by turbulence \citep{T2023a}.

\subsubsection{Lyman decrement}\label{s4-2-2}
We examine the Lyman decrement, which serves as a diagnostic of the plasma column
density \citep{C2020a}. The line ratio Ly$\beta$/Ly$\alpha$ increases first as the
column density increases, because the multiple resonance scatterings effectively reduce
the escape probability of Ly$\alpha$ photons more than Ly$\beta$ photons with a smaller
$A$ value. However, as the column density grows even larger, the Ly$\beta$ photons
suffer degradation and the ratio turns to a decreasing trend towards 0. We use the sum of
the fine-structure doublet both for Ly$\alpha$ and Ly$\beta$.

We first performed a \texttt{Cloudy} simulation of a shell with a uniform density of
$10^{10}~\mathrm{cm}^{-3}$ and a varying column density from $10^{20}~\mathrm{cm}^{-2}$
to $10^{24}~\mathrm{cm}^{-2}$ in steps of 0.2 dex. The shell was photo-ionised by a
point-like source at the centre, emitting isotropic and unpolarised radiation with a
flat SED. The ionisation parameter at the shell surface towards the incident source was
fixed at $10^{3.5}$~erg~s$^{-1}$~cm. We next performed a \texttt{SKIRT} simulation with
the same setup by using the spatial distribution of the temperature and ion density
obtained with the \texttt{Cloudy} simulation.

\begin{figure}
 \begin{center}
  \includegraphics[width=1.0\columnwidth,clip]{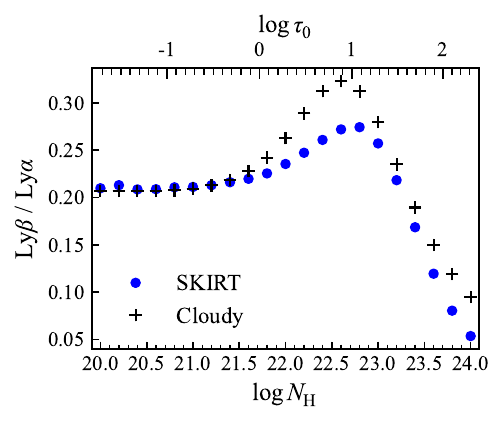}
 \end{center}
 \caption{Fe Ly$\beta$/Ly$\alpha$ ratio calculated with \texttt{SKIRT} (blue circle) and
 \texttt{Cloudy} (black plus). The top axis shows the Ly$\alpha_1$ optical depth
 ($\tau_0$), while the bottom axis shows the equivalent column density
 ($N_{\mathrm{H}}$). The offset at $\tau_0\sim1$ is presumably due to the different
 treatment of the line-photon escape or the spherical geometry.}
 \label{f08}
\end{figure}

In Fig.~\ref{f08}, we compare the Lyman decrement in the reprocessed component of the
synthesised spectrum between the two simulations. The absolute values at the peak around
$\tau_0\sim1$ differ by 16\%. This level of discrepancy is not surprising, considering
the different geometrical treatments in 3D MC-RT and 1D RT calculations, together with the 
different treatment of line-photon escape described in Sect.~\ref{s4-2-1}.
Limitations of the escape probability approximation in estimating line intensities have also been reported
in studies based on the accelerated lambda iteration (ALI) solver
\citep{C2005a}. Nevertheless, the column density at which the Lyman decrement turns into
an increasing or decreasing trend matches well between the two simulations.

\subsubsection{Formation of P Cygni profile}\label{s4-2-3}
We finally verify the \texttt{SKIRT} capability to model a P Cygni profile, which is
one of the motivations of this study. For comparison, we utilised a code\footnote{The
code is available at \url{https://github.com/unoebauer/public-astro-tools/tree/master/pcygni_profile}. A use case
is presented in \citet{N2019}.} developed for supernova ejecta flows. An analytical
expression of the profile of a line is given for a homologously expanding sphere under
the Sobolev condition with a large velocity gradient \citep{J1990}. An exponential
optical-depth profile is also assumed \citep{T2011}.

\begin{figure}
 \begin{center}
  \includegraphics[width=1.0\columnwidth,clip]{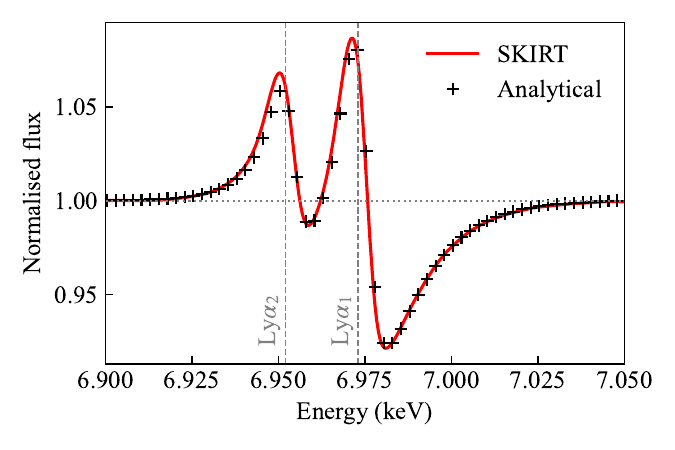}
 \end{center}
 \caption{P Cygni profile of the Fe Ly$\alpha$ doublet in a homologously expanding
 sphere calculated using \texttt{SKIRT} (red solid curve) and an analytical solution
 \citep{J1990} (black plus). For display purposes, the spectra are convolved with a 5~eV
 FWHM Gaussian and binned to 0.5 eV to be comparable to the XRISM resolution. The
 horizontal dotted line shows the incident continuum level and the vertical dashed lines
 mark the rest-frame energies of the two lines. The parameters of the code are set as
 follows: time since explosion (3000~s), maximum ejecta velocity ($0.01c$), photospheric
 velocity ($5 \times 10^{-4} c$), line optical depth (0.1), and a parameter for the
 density stratification ($5.0 \times 10^{7}$~cm~s$^{-1}$).}
\label{f09}
\end{figure}

We ran the \texttt{SKIRT} simulation with the same velocity and density structure with
the code for the Fe Ly$\alpha_{1,2}$ doublet. The maximum velocity is 0.01$c$ and the
line optical depth is $\tau_0 = 0.1$. An excellent agreement was found between the two
(Fig.~\ref{f09}) for the P Cygni profile consisting of the scattered component with
emission lines centred close to the rest frame energy and the transmitted component
with blue-shifted absorption lines. This validates the \texttt{SKIRT} calculation of
resonance scattering under a complex velocity structure.

\section{Demonstration}\label{s5}
\begin{figure}
 \begin{center}
  \includegraphics[width=0.95\columnwidth,clip]{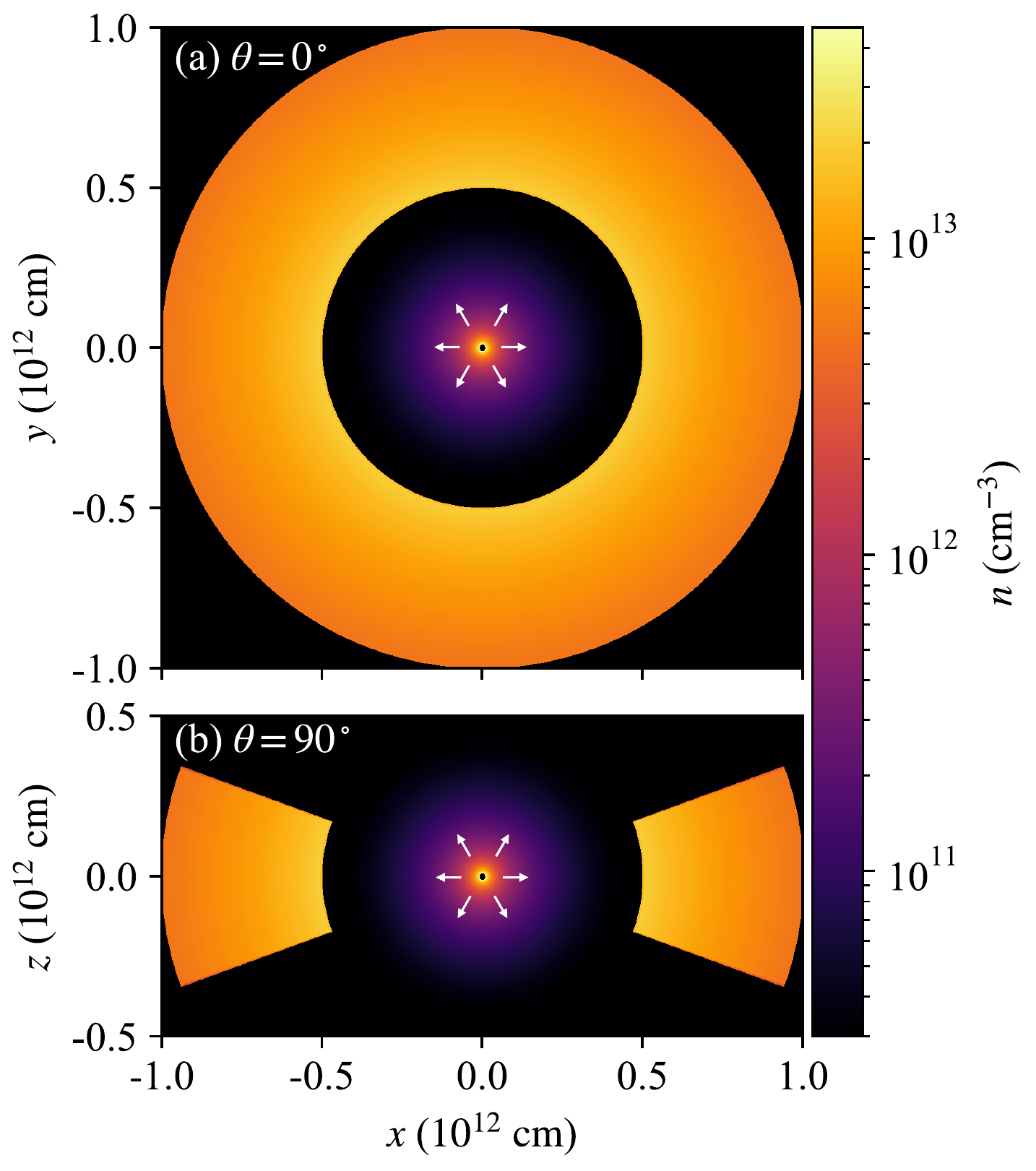}
 \end{center}
 \caption{Density distribution in the (a) face-on and (b) edge-on
 view. The radially outflowing wind is at the centre. The outer disc is static.}
 \label{f10}
\end{figure}

\begin{figure}
 \begin{center}
  \includegraphics[width=1.0\columnwidth,clip]{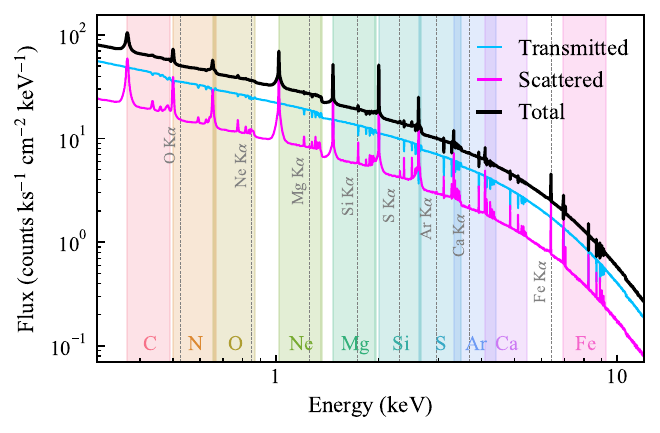}
 \end{center}
 \caption{Spectra synthesised with \texttt{SKIRT} for the model shown in Fig.~\ref{f10},
 seen from a viewing angle of 60$^{\circ}$. The transmitted, scattered, and total
 spectra are shown as cyan, magenta, and black curves, respectively. For display
 purposes, the spectra are convolved with a 5~eV FWHM Gaussian and binned to 0.5 eV to
 be comparable to the XRISM resolution. The vertical bands indicate the energy range of
 the Lyman-series lines (Ly$\alpha$ to the series limit) for selected elements in
 different colours. The dashed vertical lines mark the energy of the K$\alpha$
 fluorescent line of neutral materials.}
 \label{f11}
\end{figure}

We have verified individual processes and some RT effects in idealised setups in
Sect.~\ref{s4}. We now make a demonstration for a more realistic setup for photo-ionised
plasma around a compact object. We assume a simple geometry of a low-mass X-ray binary
consisting of a cold outer disc and a hot outflowing wind (Fig.~\ref{f10}) and
synthesise the Fe Ly$\alpha$ lines in different viewing angles.

The outer disc has a half-opening angle of 20$^{\circ}$, with an inner and outer radius of
$5\times10^{11}$ and $10^{12}$~cm, respectively. The density profile follows a power-law
of the radius with an index $-2$. It is composed of a neutral gas of a 10$^{4}$~K
temperature of the solar abundances \citep{A1989}, but with minor elements with the abundance
of less than $10^{-6}$ being omitted. 
The column density along the mid-plane is $10^{25}$~cm$^{-2}$. The outer disc is the
production site of the fluorescent line in this simulation.

The spherical wind has an inner and outer radius of $10^{10}$ and $5\times10^{11}$~cm,
respectively. The density profile follows a radial power-law with index $-2$. The
total column density is $10^{24}$~cm$^{-2}$. It consists of H-like ions of elements with
the solar abundances above $10^{-6}$, and electrons with the same total column density. 
The charge state distribution of each element and
the temperature were obtained with the \texttt{Cloudy} simulation for the same setup
with an ionisation parameter of $10^{3}$~$\mathrm{erg\ s^{-1}\ cm}$ at the inner disc radius.
The wind has a radially outward motion with a fixed velocity of $1 \times
10^{2}$~km~s$^{-1}$. The wind is the production site of the Lyman-series lines.

The incident emission is point-like, but its angular distribution is anisotropic. We
assumed the following form, which represents the emission from an inner accretion disc
\citep{N1987}.
\begin{align}
 \label{e18}
 L(\theta) \propto
 \begin{cases}
  \cos\theta \left( 2\cos\theta + 1 \right), & 0 \le \theta \le \frac{\pi}{2}, \\
  \cos\theta \left( 2\cos\theta - 1 \right), & \frac{\pi}{2} \le \theta \le \pi.
 \end{cases}
\end{align}
where $\theta$ is the inclination angle from the disc axis. Source photons are sampled from a tabulated
multi-temperature blackbody spectrum generated using the \texttt{diskbb} model in \texttt{XSPEC}
\citep{A1996}. The temperature at the inner disc radius is 2~keV, with a luminosity of $5\times10^{36}$~erg~s$^{-1}$ 
over 0.1--20~keV, corresponding to $\xi=10^{3}$~$\mathrm{erg\ s^{-1}\ cm}$ at the inner 
radius of the wind. Source photons are unpolarised.

\begin{figure}
 \begin{center}
  \includegraphics[width=1.0\columnwidth,clip]{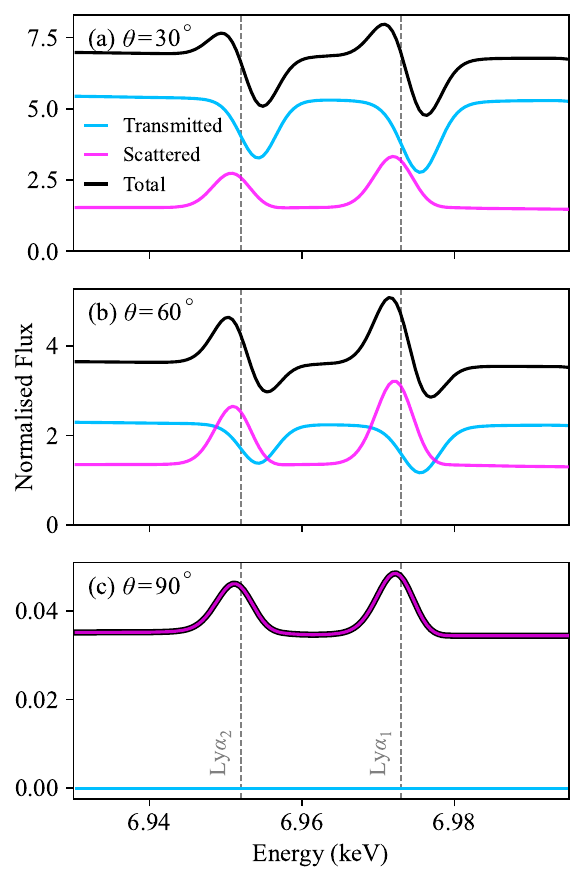}
 \end{center}
 \caption{Close-up view of the synthesised spectra of the Fe Ly$\alpha$ doublet with
 \texttt{SKIRT} at different viewing angles: (a) $\theta=30^{\circ}$, (b)
 $\theta=60^{\circ}$ and (c) $\theta=90^{\circ}$. The transmitted, scattered, and total
 spectra are shown as cyan, magenta, and black curves, respectively. For display
 purposes, the spectra are convolved with a 5~eV FWHM Gaussian and binned to 0.5 eV to
 be comparable to the XRISM resolution. The vertical dashed lines mark the rest-frame
 energies of the two lines.}
 \label{f12}
\end{figure}

\begin{figure}
 \begin{center}
  \includegraphics[width=0.9\columnwidth,clip]{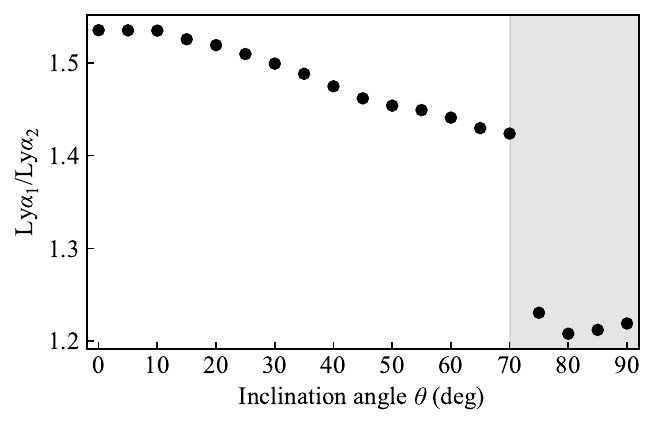}
 \end{center}
 \caption{Fe Ly$\alpha_1$/Ly$\alpha_2$ ratio as a function of the viewing angle
 calculated with \texttt{SKIRT}. Shielding by the outer disc appears at $\theta > 70^{\circ}$.}
 \label{f13}
\end{figure}

We ran the \texttt{SKIRT} simulation with $5\times10^{10}$ photon packets on a
grid with 1000 radial and 500 polar-angle cells, and synthesised the spectra
seen from different viewing angles. An example is shown for $\theta=60^{\circ}$ (Fig.~\ref{f11}), where the
outer disc lies out of the sight line. The transmitted and scattered components are
separately obtained. In the former, we observe continuum attenuation by electron
scattering, together with absorption lines and edges by H-like ions. In the latter, we
observe emission lines by H-like ions as well as fluorescence lines by neutral atoms.

A close-up view of the Ly$\alpha$ line profile is shown in Fig.~\ref{f12} at three
different viewing angles. The absorption feature imprinted on the transmitted component
is blue-shifted, while the emission feature in the scattered component is close to the
rest energies. When they are added, the P Cygni profile is formed for
$\theta=30^{\circ}$ and $60^{\circ}$. The fraction of the scattered component increases with
increasing $\theta$ due to the shielding by the outer disc coupled with the anisotropic
emission of the incident source. At the edge-on view ($\theta=90^{\circ}$), the
transmitted component is completely lost and only the emission line is observed.

We plotted the ratio of the fine-structure doublet Ly$\alpha_1$/Ly$\alpha_2$ as a
function of the viewing angle ($\theta$) in Fig.~\ref{f13}. The line intensity was
evaluated by subtracting the continuum level in the scattered component. At
$\theta=0^{\circ}$, the ratio is $\sim$1.53, which is smaller than the statistical
weight ratio of 2 for the doublet. This is a known RT effect at high line optical depth,
because multiple resonance scatterings effectively reduce the escape probability of
Ly$\alpha_1$ photons more than Ly$\alpha_2$ photons \citep{G2025a}. As $\theta$
increases, this ratio decreases because of scattering phase function effects. For
unpolarised incident emission ($I > 0$, $Q=0$, $U=0$), the scattered emission is
proportional to $M_{11} I$, using the (1,1) element of the M\"{u}ller matrix. In
Eq.~(\ref{e10}), $M_{11}(\theta)=\frac{7+3\cos^{2}{\theta}}{16}$ for Ly$\alpha_1$ and
$M_{11}(\theta)=\frac{1}{2}$ for Ly$\alpha_2$. Therefore, the ratio
\begin{equation}
 \label{e19}
  \frac{\mathrm{Ly}\alpha_1}{\mathrm{Ly}\alpha_2} \propto \frac{7+3\cos^{2}{\theta}}{8}
\end{equation}
decreases by a factor of $\frac{7}{10}$ as $\theta$ increases from 0$^{\circ}$ to
90$^{\circ}$. This change is convolved with the angular distribution of the incident
emission (Eq.~(\ref{e18})). At $\theta>70^{\circ}$, only the scattered component is
observed, thus this effect is fully seen with little dilution by the transmitted
component.

\section{Discussion}\label{s7}
The present work establishes \texttt{SKIRT} as a promising tool for interpreting
high-resolution X-ray spectra and polarisation, opening a new avenue for constraining
the geometry and kinematics of photo-ionised plasmas in the era of XRISM, IXPE, and
future observatories such as NewAthena \citep{C2025b}. Nevertheless, it has some
limitations, and we discuss four of them along with prospects for improvement.

First, only Lyman-series lines of H-like ions are implemented in this work. The
fluorescence lines from neutral atoms have been implemented in \citet{M2023b,
M2024}. Our goal is to implement the major useful lines across all ionisation stages,
and we have now covered both ends. We will extend the coverage in both directions:
from H-like to He-like and lower ionisation stages, and from neutral to higher
ionisation stages. Some development is already underway with improved atomic models.

Second, as shown throughout the paper, we relied on 1D two-stream RT codes to set up
\texttt{SKIRT} simulations for the spatial distributions of temperature and charge
states. Achieving all these calculations within \texttt{SKIRT}, or in any MC-RT codes,
would require enormous effort to implement all the atomic and chemical processes, both
collisional and radiative. This has in fact taken many decades for \texttt{Cloudy},
\texttt{SPEX}, and \texttt{XSTAR}. For the time being, it seems sensible to continue
using a combination of the two approaches. This does not completely limit the 3D
capability of \texttt{SKIRT}, however, as we can map the results of 1D RT calculations
from two-stream codes in different directions onto a 3D setup.

Third, we do not treat the polarisation of cascade photons. \cite{K1967} experimentally 
observed some linear-polarisation correlation among successive photons emitted in an atomic 
cascade. A rigorous treatment of such effects would require tracking the magnetic sublevel 
populations and coherences throughout the cascade, which is beyond the scope of the present 
implementation. However, we do not expect this approximation to significantly affect the 
predicted line profiles, because the line flux is dominated by RS without photon degradation,
for which we implement the appropriate polarisation treatment following \cite{H1947} in Sect.~\ref{s3-1-3}.

Finally, this work focuses only on photo-ionised plasma around compact objects.
Resonance scattering can also take place in collisionally ionised (CI) plasmas. The
diagnostics using line ratios and polarisation presented in this paper have also been
explored in CI plasmas of stellar coronae and clusters of galaxies \citep{S2002,N2003,H2018}.
In these plasmas, collisional processes dominate the population and depopulation of each
level (Fig.~\ref{f02}); thus, a different formalism (Sect.~\ref{s2-3}) needs to be
adopted.  Once this is prescribed, many parts of the present implementation can be
utilised for CI plasmas.

\section{Summary}\label{s8}
In this work, we have implemented the Lyman-series lines of H-like ions in the MC-RT code
\texttt{SKIRT} as a first step towards a comprehensive treatment of photon--ion
interactions in the X-ray regime. Our main results can be summarised as follows:

\begin{enumerate}
 \item We developed a framework that consistently treats two key channels of Lyman
       series photons: resonance scattering and radiative recombination.
 \item The implemented microphysics, including cross-sections, branching ratios, and
       redistribution functions, have been validated against analytical solutions and
       show good agreement with the result by an independent RT code \texttt{Cloudy}.
 \item The code successfully reproduces RT effects such as the line-profile distortion
       due to frequency diffusion, the Lyman decrement, and the formation of P Cygni
       profiles in velocity fields.
 \item In three-dimensional geometries, we demonstrated that anisotropy of the radiation
       field, geometry, and velocity structure significantly affect observable features
       in the spectra, including the P Cygni profile and the Ly$\alpha$ fine-structure
       doublet ratio.
 \item These effects, which are well within reach of the XRISM capability, cannot be
       captured by conventional two-stream codes and highlight the importance of a MC
       code like \texttt{SKIRT}.
\end{enumerate}

\begin{acknowledgements}
 We thank Anton Krieger for refereeing this paper and providing constructive comments, 
 which greatly improved the manuscript.
 N.\,S. acknowledges the hospitality by colleagues at Ghent University,
 ESTEC and SRON, where he stayed throughout the course of this work. This research made
 use of the JAXA's high-performance computing system JSS3. N.\,S. acknowledges support
 from the JSPS Core-to-Core Program (grant number: JPJSCCA20220002), the Foundation for
 Promotion of Astronomy, and the JSPS KAKENHI Grant Number 26H02075. B.\,V.\
 acknowledges support through the European Space Agency (ESA) Research Fellowship
 Programme in Space Science.
\end{acknowledgements}

\bibliographystyle{aa}
\bibliography{main}
\end{document}